\documentclass[aps,showpacs,preprintnumbers,amsmath,amssymb,superscriptaddress,floatfix,a4paper,nofootinbib,11pt,floatfix,nofootinbib,onecolumn]{revtex4}
\usepackage{graphicx}
\usepackage{bm}
\usepackage{rotating}
\usepackage{array}
\usepackage{xcolor}
\usepackage{amsmath,amssymb}
\usepackage{mathrsfs}
\usepackage{graphicx}
\usepackage{color}
\usepackage{subfigure}
\usepackage{fancyhdr}
\usepackage{multirow}
\usepackage{float}
\usepackage{epsfig}
\usepackage{amsfonts}
\usepackage{bm}
\usepackage{multirow}
\usepackage{nicefrac}
\newcommand{\ba}{\begin{eqnarray}}
\newcommand{\ea}{\end{eqnarray}}

\usepackage{bbm}
\newcommand{\be}{\begin{eqnarray}}
\newcommand{\ee}{\end{eqnarray}}

\begin{document}

\title{Nonminimally-coupled warm Higgs inflation: Metric vs. Palatini Formulations}

\author{Thammarong Eadkhong} 
\email{thammarong.ea@mail.wu.ac.th}
\affiliation{School of Science, Walailak University, Thasala, Nakhon Si Thammarat, 80160, Thailand}

\author{Punsiri Dam-O} 
\email{dpunsiri@mail.wu.ac.th}
\affiliation{School of Science, Walailak University, Thasala, Nakhon Si Thammarat, 80160, Thailand}

\author{Phongpichit Channuie} 
\email{phongpichit.ch@mail.wu.ac.th}
\affiliation{School of Science, Walailak University, Thasala, Nakhon Si Thammarat, 80160, Thailand}
\affiliation{College of Graduate Studies, Walailak University, Thasala, \\Nakhon Si Thammarat, 80160, Thailand}

\author{Davood Momeni} 
\email{dmomeni@nvcc.edu}
\affiliation{Northern Virginia Community College, 8333 Little River Turnpike, Annandale, VA 22003}

\begin{abstract}	

In this work, we study the non-minimally-coupled Higgs model in the context of warm inflation scenario on both metric and Palatini approaches. We particularly consider a dissipation parameter of the form $\Gamma=C_{T}T$ with $C_{T}$ being a coupling parameter and focus only on the strong regime of the interaction between inflaton and radiation fluid. We compute all relevant cosmological parameters and constrain the models using the observational Planck 2018 data. We discover that the $n_s$ and $r$ values are consistent with the observational bounds. Having used the observational data, we constrain a relation between $\xi$ and $\lambda$ for the non-minimally-coupled warm Higgs inflation in both metric and Palatini cases. To produce $n_s$ and $r$ in agreement with observation, we find that their values are two orders of magnitude higher than those of the usual (cold) non-minimally-coupled Higgs inflation.
	 
\end{abstract}

%\pacs{Valid PACS appear here}

\maketitle

%%%%%%%%%%%%%%%%%%%%%%%%%%%%%%%%%%%%%%%%

%%%%%%%%%%%%%%%%%%%%%%%
\section{Introduction}
%%%%%%%%%%%%%%%%%%%%%%%

Inflationary cosmology is a widely accepted framework for explaining the exponentially rapid expansion of the early universe. The flatness and homogeneity/unwanted relics problems can be solved using such paradigm which provides a mechanism to generate the inhomogeneities in the cosmic microwave background radiation (CMBR) \cite{Starobinsky:1980te,Sato:1980yn,Guth:1980zm,Linde:1981mu,Albrecht:1982wi}. In a standard fashion of slow-roll (cold) inflation, the universe experiences an exponential expansion, during which density perturbations are created by quantum fluctuations of the inflation field, followed by the reheating stage.  However, the standard inflationary model is plagued by several challenges, such as the graceful exit problem. To address these issues, several alternative models of inflation have been proposed, including warm inflation. This is a combination of the exponential accelerating expansion phase and the reheating. Warm inflation has at the moment become a growing area of research, with the potential to provide new insights into the physics of the early universe.

Warm inflation is an alternative version of standard inflation that takes into account the effects of dissipation and thermal fluctuations on the inflationary process. In warm inflation scenario, the scalar field responsible for driving inflation, is coupled to a thermal bath and transfers energy to radiation during inflation, thus maintaining a non-zero temperature. Warm inflation was first proposed by Berera and Fang  \cite{Berera:1995wh}. Since then, numerous studies have been carried out to study the dynamics and predictions of warm inflation. One of the main advantages of warm inflation is that it provides a natural solution to the graceful exit problem, as the inflaton can gradually decay into the thermal bath, leading to a smooth transition from inflation to the hot big bang era.

The predictions of warm inflation have been studied both analytically and numerically. Some of the most notable works in this field include Berera {\it et. al.} \cite{Taylor:2000ze,Berera:1996fm,Berera:1999ws,Berera:2008ar,Berera:1998gx,BasteroGil:2009ec}, Graham and I.~G.~Moss \cite{Graham2009}, Bastero-Gil {\it et.\,al.} \cite{Bastero-Gil:2018uep} and Zhang \cite{Zhang:2009ge}. These studies have shown that warm inflation can produce a sufficient number of e-folds, consistent with the observed CMB temperature fluctuations, and that it can lead to a broad spectrum of curvature perturbations. There have also been several studies comparing the predictions of warm inflation with those of the standard inflationary model and other alternative models of inflation. For example, Kamali \cite{Kamali:2018ylz} compared warm inflation with the Higgs inflation model and found that warm inflation can produce a smaller tensor-to-scalar ratio, which is more in line with the current observations. Similarly, the authors of \cite{Bartrum:2013fia} showed that even when dissipative effects are still small compared to Hubble damping, the amplitude of scalar curvature fluctuations can be significantly enhanced, whereas tensor perturbations are generically unaffected due to their weak coupling to matter fields.

Warm Higgs inflation is recently investigated in several publications, by emphasizing on different aspects of theory.  In \cite{Motaharfar:2018mni}, the Galileon scalar field dissipation formalism is proposed via  its kinetic energy. The radiation fluid throughout inflation emerges  and it has been shown that  in this scenario, the universe smoothly can  enter into a radiation dominated era without the reheating phase. Different regimes for temperature are investigated and the backreaction of radiation on the power spectrum is calculated.  In \cite{Bastero-Gil:2018uep}, the warm Little Inflation scenario proposed as a quantum field theoretical realization of the  warm inflation. Different potential terms are used including Higgs model and chaotic potential. Keeping in mind 50-60 e-folds of inflation, by introducing a viable thermal correction to the inflaton potential term, the  primordial spectrum of different modes of perturbations and the tensor-to-scalar ratio calculated in light of the Planck data. The motivation of our study is that we provided a constraint between two parameters of the potential term.

In the context of warm inflation, it was also found that recent studies in many different theories were proposed. For instance, the authors of Ref.\cite{Dymnikova:2000gnk} conducted a possible realization of warm inflation owing to a inflaton field self-interaction. Additionally, models of minimal and non-minimal coupling to gravity were investigated in Refs.\cite{Panotopoulos:2015qwa,Benetti:2016jhf,Motaharfar:2018mni,Graef:2018ulg,Arya:2018sgw,Kamali:2018ylz}. Recently, warm scenarion of the Higgs-Starobinsky (HS) model was conducted \cite{Samart:2021eph}. The model includes a non-minimally coupled scenario with quantum-corrected self-interacting potential in the context of warm inflation \cite{Samart:2021hgt}. An investigation of warm inflationary models in the context of a general scalar-tensor theory of gravity has been made in Ref.\cite{Amake:2021bee}. Recent review on warm inflation has been recently discussed in Ref.\cite{Kamali:2023lzq}.

The physical motivations of the present work is devoted to a comparative analysis of the dynamics of non-minimally coupled scalar fields in Metric and Palatini formulations. It will be shown that the two formalisms generally yield different answers for both metric tensor and scalar fields provided that non-minimal coupling of the scalar field to curvature scalar does not vanish. Investigation both formalisms helps ensure the theoretical consistency of the gravitational theories. By investigating different formalisms, one can uncover potential limitations or inconsistencies in the assumptions made in each formalism, or even may lead to distinct predictions. This aids in refining and extending our understanding of gravity and its mathematical framework. Investigating both formalisms enables us to identify potential observational or experimental tests that could distinguish between the predictions of these formalisms and is crucial for testing the validity of different theories.

The plan of the work is structured as follows: In Sec.\ref{C2}, we review a formulation of non-minimally-coupled Higgs inflation considering both metric and Palatini approaches. In Sec.\ref{C3}, we provide the basic evolution equations for the inflaton and the radiation fields and define the slow roll parameters and conditions. We also describe the primordial power spectrum for warm inflation and the form of the dissipation coefficient. In Sec.\ref{C4}, we present the models of nonminimally-coupled Higgs inflation and compute all relevant cosmological parameters. We then constrain our models using the the observational (Planck 2018) data in Sec.\ref{C5}. Finally, we summarize the present work and outline the conclusion.

%%%%%%%%%%%%%%%%%%%%%%%
\section{A Review on Metric vs. Palatini Formulations of nonminimally-coupled Higgs inflation}\label{C2}
%%%%%%%%%%%%%%%%%%%%%%%
Models in which the Higgs field is non-minimally coupled to gravity lead to successful inflation and produce the spectrum of primordial fluctuations in good agreement with the observational data. Here we consider the theory composed of the Standard Model Higgs doublet $H_J$ with the non-minimal coupling to gravity in the Jordan (J) frame:
\begin{eqnarray}\label{starting action}
    S_{J}=\int d^{4}x\sqrt{-g_J}\left[\frac{M_{p}^2}{2}\left(1+2\xi\frac{H_{J}^{\dagger} H_J}{M_{p}^2}\right)R_J+g_J^{\mu\nu}(D_\mu H_J)^\dag (D_\nu H_J)-\lambda(H_J^\dagger H_J)^2\right],
\end{eqnarray}
where $M_p$ is the Planck mass, $\xi$ is a coupling constant, $R_J$ is the Ricci scalar, and $H$ is the Higgs field with $\lambda$ being the self-coupling of the Higgs doublet. Note that the mass term of the Higgs doublet is neglected throughout this paper because it is irrelevant during inflation.

As was known the metric formalism is considered as a standard gravitational method, however, one can study gravity adopting the Palatini approach leading to different phenomenological consequences in a theory with a non-minimal coupling to gravity. The differences of them are explicit and easily understandable in the so-called Einstein (E) frame where the non-minimal coupling is removed from the theory by taking a conformal redefinition of the metric
\begin{eqnarray}\label{conformal}
   g_{\mu\nu}\rightarrow \Omega^2 g_{J,\mu\nu},\qquad \Omega^2 = 1+2\xi\frac{H_{J}^{\dagger}H_{J}}{M_{p}^2}.
\end{eqnarray}
Using the metric redefinition, the connection is also transformed in the metric formalism since it is given by the Levi-Civita connection:
\begin{eqnarray}
 \Gamma^\rho_{\mu\nu}(g)=\frac{1}{2}g^{\rho\lambda}\Big(\partial_\mu g_{\nu\lambda}+\partial_\nu g_{\mu\lambda}-\partial_\lambda g_{\mu\nu}\Big).
\end{eqnarray}
It is noticed that the connection is left unaffected in the Palatini formalism because it is treated as an independent variable as well as the metric. Thus, the Ricci scalar transforms differently depending on the underlying gravitational formulations as~\cite{Tenkanen:2020dge}
\begin{eqnarray}
    \sqrt{-g_J}\Omega^2\,R_J=\sqrt{-g}(R+6\kappa\Omega g^{\mu\nu}\nabla_\mu\nabla_\nu \Omega^{-1}),
\end{eqnarray}
where $\kappa=1$ and $\kappa=0$ correspond to the metric and the Palatini formalism, respectively. The Einstein frame expression can then be obtained after the rescaling of the metric:
\begin{eqnarray}\label{fulleinstein}
    S=\int d^{4}x\sqrt{-g}\left[\frac{M_{p}^2}{2}R+3\kappa M_{p}^2\Omega g^{\mu\nu}\nabla_\mu\nabla_\nu\Omega^{-1}+\frac{1}{\Omega^2}g^{\mu\nu}(D_\mu H_J)^\dag (D_\nu H_J)-\frac{\lambda}{\Omega^4}(H_{J}^{\dagger}H_J)^2\right].
\end{eqnarray}
In the Einstein frame, the connection is not directly coupled to the Higgs field $H_J$ and the gravity sector is just the Einstein-Hilbert form. In this case, the Euler-Lagrange constraint in the Palatini formalism restricts the connection to the Levi-Civita one, and the two approaches become equivalent, up to the explicit difference in the $\kappa$ term~\cite{Tenkanen:2020dge}.

Let us next review phenomenological aspects of the metric-Higgs inflation~\cite{Bezrukov:2007ep} and the Palatini-Higgs inflation~\cite{Bauer:2008zj,Jinno:2019und,Takahashi:2018brt,Mikura:2021clt}. In this subsection, we neglect the gauge sector for simplicity. In the inflationary fashion, we usually consider the unitary gauge in which the Higgs doublet is described by a real scalar field $\phi_J(x)$ as $H_{J}^T(x)=(0,\phi(x)/\sqrt{2})$. Therefore, the action in Eq.~\eqref{fulleinstein} becomes
\begin{eqnarray}\label{E-frame unitary}
    S_E=\int d^{4}x\sqrt{-g}\Bigg(\frac{M_{p}^2}{2}R-\frac{1+\xi\frac{\phi^2}{M_{p}^2}+6\kappa\xi^2\frac{\phi^2}{M_{p}^2}}{2\left(1+\xi\frac{\phi^2}{M_{p}^2}\right)^2}g^{\mu\nu}\partial_\mu\phi\partial_\nu\phi-\frac{\lambda \phi^4}{4\left(1+\xi\frac{\phi^2}{M_{p}^2}\right)^2}\Bigg),
\end{eqnarray}
where $\kappa=1$ for the metric-Higgs and $\kappa=0$ for the Palatini-Higgs inflation. The non-trivial kinetic term can be canonically normalized by introducing the field $\psi$ defined through
\begin{eqnarray}\label{canonical transformation}
    \frac{d\psi}{d\phi}=\sqrt{\frac{1+\xi\frac{\phi^2}{M_{p}^2}+6\kappa\xi^2\frac{\phi^2}{M_{p}^2}}{\left(1+\xi\frac{\phi^2}{M_{p}^2}\right)^2}}.
\end{eqnarray}
In terms of $\psi$, the action can be rewritten as 
\begin{eqnarray}
S_E=\int d^{4}x\sqrt{-g}\Big(\frac{M_{p}^2}{2}R-\frac{1}{2}g^{\mu\nu}\partial_\mu\psi\partial_\nu\psi-U(\psi(\phi))\Big),
\end{eqnarray}
with the potential in the Einstein frame
\begin{eqnarray}
U(\psi(\phi))=\frac{\lambda \phi^4(\psi)}{4\left(1+\xi\frac{\phi^2(\psi)}{M_{p}^2}\right)^2}.
\end{eqnarray}
The change of variable can be easily integrated in the Palatini case, while an asymptotic form in the large field limit $\xi\phi^2/M_p^2\gg1$ is useful in the metric case as
\begin{eqnarray}
\mathrm{metric}\quad\phi&\simeq& \frac{M_p}{\sqrt{\xi}}\mathrm{exp}\Bigg(\sqrt{\frac{1}{6}}\frac{\psi}{M_p}\Bigg)\,,\\
\mathrm{Palatini}\quad\phi &=&\frac{M_p}{\sqrt{\xi}}\mathrm{sinh}\left(\frac{\sqrt{\xi}\psi}{M_p}\right).
\end{eqnarray}
The potential is reduced to
\begin{eqnarray}\label{metric potential}
    \mathrm{metric}:\quad U&\simeq& \frac{\lambda M_p^4}{4\xi^2}\left(1+\mathrm{exp}\left(-\sqrt{\frac{2}{3}}\frac{\psi}{M_p}\right)\right)^{-2}, \\
    \label{Palatini potential}
    \mathrm{Palatini}:\quad U&=&\frac{\lambda M_p^4}{4\xi^2}\mathrm{tanh}^4\left(\frac{\sqrt{\xi}\psi}{M_p}\right).
\end{eqnarray}
The potentials in both scenarios approach asymptotically to a constant value $U\simeq \frac{\lambda M_p^4}{4\xi^2}$ at a large field region, which is suitable for slow-roll inflation. An observed amplitude ${\cal P}_\zeta\simeq 2.2\times 10^{-9}$~\cite{Planck:2018jri} fixes the relation between $\xi$ and $\lambda$ in the metric and Palatini approaches, $\xi_{\rm met}\sim 5\times10^4\sqrt{\lambda},\,\xi_{\rm Pal}\sim 10^{10}\lambda$, respectively. The CMB normalization restricts that the coupling to gravity $\xi$ should be quite large unless the quartic coupling $\lambda$ is extremely small both in the metric and Palatini formalisms, see also models with non-minimal coupling in metric and Palatini formalisms \cite{Cheong:2021kyc}.

%%%%%%%%%%%%%%%%%%%%%%%%%%%%%%%%%
\section{Theory of warm inflation revisited}\label{C3}
%%%%%%%%%%%%%%%%%%%%%%%%%%%%%%%%%
The warm inflation dynamics is characterized by the coupled system of the background equation of motion for the inflaton field, $\psi(t)$, the evolution equation for the radiation energy density, $\rho_{r}(t)$. Considering the Einstein frame action with the flat FLRW line element, the Friedmann equation for warm inflation taks the form
\begin{eqnarray}
H^2 =\frac{1}{3\,M_p^2}\left( \rho_{\psi} + \rho_r\right)= \frac{1}{3\,M_p^2}\left( \frac{1}{2}\,\dot\psi^2 + U(\psi) + \rho_r\right)\,,\label{E1}
\end{eqnarray}
with $\dot{\psi}=d\psi/dt$ and $\rho_{r}$ being the energy density of the radiation fluid with the equation of state given by $w_{r}=1/3$. The Planck 2018 baseline plus BK15 constraint on $r$ is equivalent to an upper bound on the Hubble parameter during inflation of $H_{*}/M_{p}<2.5\times 10^{-5}\,(95 \%\,CL)$ \cite{Planck:2018jri}. The equation of motion of the homogeneous inflaton field $\phi$ during warm inflation is governed as
\begin{eqnarray}
\ddot\psi + 3H\,\dot\psi + U'(\psi) = -\Gamma\,\dot\psi\,,
\end{eqnarray}
where $U'(\psi)=dU(\psi)/d\psi$. The above relation is equivalent to the evolution equation for the inflaton energy density $\rho_\phi$ given by
\begin{eqnarray}
\dot \rho_\psi + 3 H ( \rho_\psi + p_\psi) = - \Gamma ( \rho_\psi +p_\psi) \,,
\label{rhoinf}
\end{eqnarray}
with pressure $p_\psi = \dot \psi^2/2 - U(\psi)$, and $\rho_\psi + p_\psi= \dot \psi^2$. Here the  RHS of Eq. (\ref{rhoinf}) acts as the source term. In case of radiation, we have
\begin{eqnarray}
\dot \rho_r + 4 H \rho_r  = \Gamma \dot{\psi}^2\,. \label{eomrad}
\end{eqnarray}
A condition for warm inflation requires $\rho^{1/4}_{r}>H$ in which the dissipation potentially affects both the background inflaton dynamics, and the primordial spectrum of the field fluctuations. Following Refs.\cite{Zhang:2009ge,Bastero-Gil:2011rva}, we consider the general form of the dissipative coefficient, given by
\begin{eqnarray}
\Gamma  = C_{m}\frac{T^{m}}{\psi^{m-1}}\,, \label{Q1}
\end{eqnarray}
where $m$ is an integer and $C_{m}$ is associated to the dissipative microscopic dynamics which is a measure of inflaton dissipation into radiation. Different choices of $m$ yield different physical descriptions, e.g., Refs.\cite{Zhang:2009ge,Bastero-Gil:2011rva,Bastero-Gil:2012akf}. For $m=1$, the authors of Refs.\cite{Berera:2008ar,Panotopoulos:2015qwa,Bastero-Gil:2016qru} have discussed the high temperature regime. For $m=3$, a supersymmetric scenario has been implemented \cite{Berera:2008ar,Bastero-Gil:2011rva,Bastero-Gil:2010dgy}. A minimal warm inflation was also proposed \cite{Berghaus:2019whh,Laine:2021ego,Motaharfar:2021egj}. Particularly, it was found that thermal effects suppress the tensor-to-scalar ratio $r$ significantly, and predict unique non-gaussianities. Apart from the Hubble term, the present of the extra friction term, $\Gamma$, is relevant in the warm scenario. In slow-roll regime, the equations of motion are governed by
\begin{eqnarray}
3 H ( 1 + Q ) \dot \psi &\simeq&  -U_\psi    \,,\label{eominfsl} \\
4 \rho_r  &\simeq& 3 Q\dot \psi^2\,. \label{eomradsl}
\end{eqnarray}
where the dissipative ratio $Q$ is defined as $Q=\Gamma/(3 H)$ and $Q$ is not necessarily constant. Since the coefficient $\Gamma$ depends on $\phi$ and $T$, the dissipative ratio $Q$ may increase or decrease during inflation.  The flatness of the potential $U(\psi)$ in warm inflation is measured in terms of the slow roll parameters which are defined in Ref.\cite{Hall:2003zp} given by
\begin{eqnarray}
\varepsilon &=& \frac{M_p^2}{2}\left( \frac{U'}{U}\right)^2\,,\quad \eta = M_p^2\,\frac{U''}{U}\,,\quad \beta = M_p^2\left( \frac{U'\,\Gamma'}{U\,\Gamma}\right)\,.
\label{SR-parameters}
\end{eqnarray}
Since a $\beta$ term depends on $\Gamma$ and hence disappears in standard cold inflation. In warm inflationary model, the slow roll parameters are defined as follows:
\begin{eqnarray}
\varepsilon_{H} =\frac{\varepsilon}{1+Q}\,,\quad \eta_{H} = \frac{\eta}{1+Q}\,.
\label{slowroll}
\end{eqnarray}
Inflationary phase of the universe in warm inflation takes place when the slow-roll parameters satisfy the following conditions \cite{Hall:2003zp,Taylor:2000ze,Moss:2008yb}:
\begin{eqnarray}
\varepsilon \ll 1 + Q\,,\qquad \eta \ll 1 + Q\,,\qquad \beta \ll 1 + Q\,,\label{sloe}
\end{eqnarray}
where the condition on $\beta$ ensures that the variation of $\Gamma$ with respect to $\phi$ is slow enough. Compared to the cold scenario, the power spectrum of warm inflation gets modified and it is given in Refs.\cite{Graham2009,Bastero-Gil:2018uep,Hall:2003zp,Ramos:2013nsa,BasteroGil:2009ec,Taylor:2000ze,DeOliveira:2001he,Visinelli:2016rhn}
and it takes the form:
\begin{eqnarray}
P_{\cal R}(k) = \left( \frac{H_k^2}{2\pi\dot\phi_k}\right)^2\left( 1 + 2n_k +\left(\frac{T_k}{H_k}\right)\frac{2\sqrt{3}\,\pi\,Q_k}{\sqrt{3+4\pi\,Q_k}}\right)G(Q_k)\,,
\label{spectrum}
\end{eqnarray}
where the subscript $``k"$ signifies the time when the mode of cosmological perturbations with wavenumber $``k"$ leaves the horizon during inflation and $n = 1/\big( \exp{H/T} - 1 \big)$ is the Bose-Einstein distribution function. Additionally, the function $G(Q_k)$ encodes the coupling between the inflaton and the radiation in the heat bath leading to a growing mode in the fluctuations of the inflaton field. It is originally proposed in Ref.\cite{Graham2009} and its consequent implications can be found in Refs.\cite{BasteroGil:2011xd,BasteroGil:2009ec}.

This growth factor $G(Q_k)$ is dependent on the form of $\Gamma$ and is obtained numerically. As given in Refs.\cite{Benetti:2016jhf,Bastero-Gil:2018uep}, we see that for $\Gamma \propto T$:
\be
G(Q_k)_{\rm linear} = 1+0.0185Q^{2.315}_{k}+0.335Q^{1.364}_{k}
\,. \label{ga}
\ee
In this work, we consider a linear form of $G(Q_k)$ with $Q\gg 1$. Clearly, for small $Q$, i.e., $Q\ll 1$, the growth factor does not enhance the power spectrum. It is called the weak dissipation regime. However, for large $Q$, i.e., $Q\gg 1$, the growth factor significantly enhances the power spectrum. The latter is called the strong dissipation regime. The primordial tensor fluctuations of the metric give rise to a tensor power spectrum. It is the same form as that of cold inflation given
in Ref.\cite{Bartrum:2013fia} as
\be
P_{T}(k) = \frac{16}{\pi}\Big(\frac{H_{k}}{M_{p}}\Big)^{2}
\,. \label{PT}
\ee
The ratio of the tensor to the scalar power spectrum is expressed in terms of a parameter $r$ as
\be
r = \frac{P_{T}(k)}{P_{\cal R}(k)}\,. \label{r}
\ee
As of the primordial power spectrum for all the models written in terms of $Q$,\,$\lambda$,\,and $C_{1}$, we can demonstrate how the the power spectrum does depend on the scale. The spectral index of the primordial power spectrum is defined as
\begin{eqnarray}
n_{s}-1=\frac{d \ln P_{\cal R}(k)}{d\ln (k/k_{p})}=\frac{d\ln P_{\cal R}(k)}{dQ}\frac{dQ}{dN}\frac{dN}{dx}\Bigg|_{k=k_{p}}\,,\label{ns1}
\end{eqnarray}
where $x=\ln(x/x_{p})$ and $k_{p}$ corresponds to the pivot scale. From a definition of $N$, it is rather straightforward to show that \cite{Arya:2018sgw}
\begin{eqnarray}
\frac{dN}{dx}=-\frac{1}{1-\varepsilon_{H}}\,.
\end{eqnarray}
Now we compute $r$ and $n_{s}$ using Eq.(\ref{r}) and Eq.(\ref{ns1}) for a linear form of the growing mode function $G(Q)$ given in Eq.(\ref{ga}). Note that $r$ and $n_{s}$ are approximately given in Refs. \cite{Bastero-Gil:2018uep,BasteroGil:2009ec,Benetti:2016jhf}.

%%%%%%%%%%%%%%%%%%%%%%%%%%%%%%%%%
\section{Models of nonminimally-coupled warm inflation considered}\label{C4}
%%%%%%%%%%%%%%%%%%%%%%%%%%%%%%%%

%%%%%%%%%%%%%%%%%%%%%%%%%%%%%%%%%
\subsection{Metric Formalism}
%%%%%%%%%%%%%%%%%%%%%%%%%%%%%%%%

The energy density during inflation is predominated by the potential of the inflaton field. Therefore, we can write 
\begin{eqnarray}
H^{2}=\frac{\lambda  M_p^2}{12 \xi ^2 \left(e^{-\frac{\sqrt{\frac{2}{3}} \psi }{M_p}}+1\right)^2}\,.\label{H2}
\end{eqnarray}
Using this we can express Eq.(\ref{E1}) for this model as
\begin{eqnarray}
{\dot \psi}\approx -\frac{U'(\psi)}{3 (Q+1) H}=-\frac{\sqrt{2} M_p e^{\frac{\sqrt{\frac{2}{3}} \psi }{M_p}} \sqrt{\frac{\lambda  M_p^2}{\xi ^2}}}{3 (Q+1) \left(e^{\frac{\sqrt{\frac{2}{3}} \psi }{M_p}}+1\right)^2}\,.\label{H2psi}
\end{eqnarray}
Using Eq.(\ref{H2}) and Eq.(\ref{H2psi}), we come up with the following expression:
\begin{eqnarray}
\frac{H^{2}}{2\pi {\dot \psi}}=-\frac{(Q+1) e^{\frac{\sqrt{\frac{2}{3}} \psi }{M_p}} \sqrt{\frac{\lambda  M_p^2}{\xi ^2}}}{8 \sqrt{2} \pi  M_p}\,.\label{H2psid}
\end{eqnarray}
On substituting $Q=\Gamma/3H=C_{T}T/3H$ in the energy density of radiation given in Eq.(\ref{eomradsl}), we obtain the temperature of the thermal bath as
\begin{eqnarray}
T=\frac{1}{6^{1/4}}\left(\frac{\lambda  Q M_p^4 e^{\frac{2 \sqrt{\frac{2}{3}} \psi }{M_p}}}{C_{r} \xi ^2 (Q+1)^2 \left(e^{\frac{\sqrt{\frac{2}{3}} \psi }{M_p}}+1\right)^4}\right)^{1/4}\,.\label{Tex}
\end{eqnarray}
Dividing the above relation with $H$, we find
\begin{eqnarray}
\frac{T}{H}=2^{3/4}\,3^{1/4}\left(\frac{\lambda  M_p^2}{\xi ^2 \left(e^{-\frac{\sqrt{\frac{2}{3}} \psi }{M_p}}+1\right)^2}\right)^{-1/2}\left(\frac{\lambda  Q M_p^4 e^{\frac{2 \sqrt{\frac{2}{3}} \psi }{M_p}}}{\text{Cr} \xi ^2 (Q+1)^2 \left(e^{\frac{\sqrt{\frac{2}{3}} \psi }{M_p}}+1\right)^4}\right)^{1/4}\,.\label{TH}
\end{eqnarray}
The dissipation parameter is defined as $Q=\Gamma/3H=C_{T}T/3H$. In this model of warm inflation, we have $3H$ considered $\Gamma = C_{T}T$. On substituting this form of $\Gamma$ we get $T=3HQ/C_{T}$. We equate this with Eq.(\ref{Tex}) to obtain
\begin{eqnarray}
e^{\frac{\sqrt{\frac{2}{3}} \psi }{M_p}}\approx \frac{2 \sqrt{\frac{2}{3}} C_{t}^2 \xi }{3 \sqrt{C_{r}} \sqrt{\lambda } Q^{5/2}}\quad\rightarrow\quad\psi=\sqrt{\frac{3}{2}} M_{p} \log \left(\frac{2 \sqrt{\frac{2}{3}} C_{T}^2 \xi }{3 \sqrt{C_{r}} \sqrt{\lambda } Q^{5/2}}\right)\,.\label{phiex}
\end{eqnarray}
On substituting Eq.(\ref{phiex}) in Eqs.(\ref{Tex}) and (\ref{H2psid}), we can express $P_{R}(k)$ in terms of variables
$\xi,\,\lambda,\,Q$ and $C_T$. Also, from its definition in Eq.(\ref{slowroll}), the slow roll parameters can be written
\begin{eqnarray}
\varepsilon_{H}&=&\frac{4 e^{-\frac{2 \sqrt{\frac{2}{3}} \psi }{M_p}}}{3 (Q+1) \left(e^{-\frac{\sqrt{\frac{2}{3}} \psi }{M_p}}+1\right)^2}=\frac{9 \lambda  Q^5 C_r}{2 \xi ^2 (Q+1) C_t^4 \left(\frac{3 \sqrt{\frac{3}{2}} \sqrt{\lambda } Q^{5/2} \sqrt{C_r}}{2 \xi  C_T^2}+1\right)^2}\,,\\\eta_{H}&=&\frac{4 \xi ^2 \left(e^{-\frac{\sqrt{\frac{2}{3}} \psi }{M_p}}+1\right)^2}{\lambda  (Q+1) M_p^2}\Bigg(\frac{\lambda  M_p^2 e^{-\frac{2 \sqrt{\frac{2}{3}} \psi }{M_p}}}{\xi ^2 \left(e^{-\frac{\sqrt{\frac{2}{3}} \psi }{M_p}}+1\right)^4}-\frac{\lambda  M_p^2 e^{-\frac{\sqrt{\frac{2}{3}} \psi }{M_p}}}{3 \xi ^2 \left(e^{-\frac{\sqrt{\frac{2}{3}} \psi }{M_p}}+1\right)^3}\Bigg)\nonumber\\&=&\frac{16 \left(9 C_{r} \lambda  Q^5-\sqrt{6} \sqrt{C_{r}} C_{T}^2 \sqrt{\lambda } \xi  Q^{5/2}\right)}{(Q+1) \left(3 \sqrt{6} \sqrt{C_{r}} \sqrt{\lambda } Q^{5/2}+4 C_{T}^2 \xi \right)^2}\,.
\end{eqnarray}
Using Eq.(\ref{H2}), the tensor power spectrum for this model is evaluated and we can use Eq.(\ref{phiex}) and express $P_{T}(k)$ in terms of model parameters
\begin{eqnarray}
P_{T}(k)=\frac{16}{\pi}\Big(\frac{H}{M_{p}}\Big)^{2}=\frac{4 \lambda }{3 \pi  \xi ^2 \left(e^{-\frac{\sqrt{\frac{2}{3}} \psi }{M_p}}+1\right)^2}=\frac{4 \lambda }{3 \pi  \xi ^2 \left(\frac{3 \sqrt{\frac{3}{2}} \sqrt{\lambda } Q^{5/2} \sqrt{C_r}}{2 \xi  C_T^2}+1\right)^2}\,.
\end{eqnarray}
In this subsection, we will evaluate how the dissipation parameter, $Q$, evolves with the number of efolds, $N$. We differentiate Eq.(\ref{phiex}) w.r.t $N$ and then again write $d\psi/dN=-{\dot\psi}/H$. By using Eqs. (\ref{H2}), (\ref{H2psi}) and (\ref{phiex}), we obtain
\begin{eqnarray}
\frac{dQ}{dN}=-\frac{2 \sqrt{6} \sqrt{\lambda} Q^{5/2} \sqrt{C_r}}{5 \xi  C_T^2}\,,
\end{eqnarray}
where we have assumed a large field approximation $\xi\phi^{2}/M_{p}^{2}\gg 1$ or $\psi\gg\sqrt{3/2}M_{p}$.
\begin{figure}[!h]	
	\includegraphics[width=8cm]{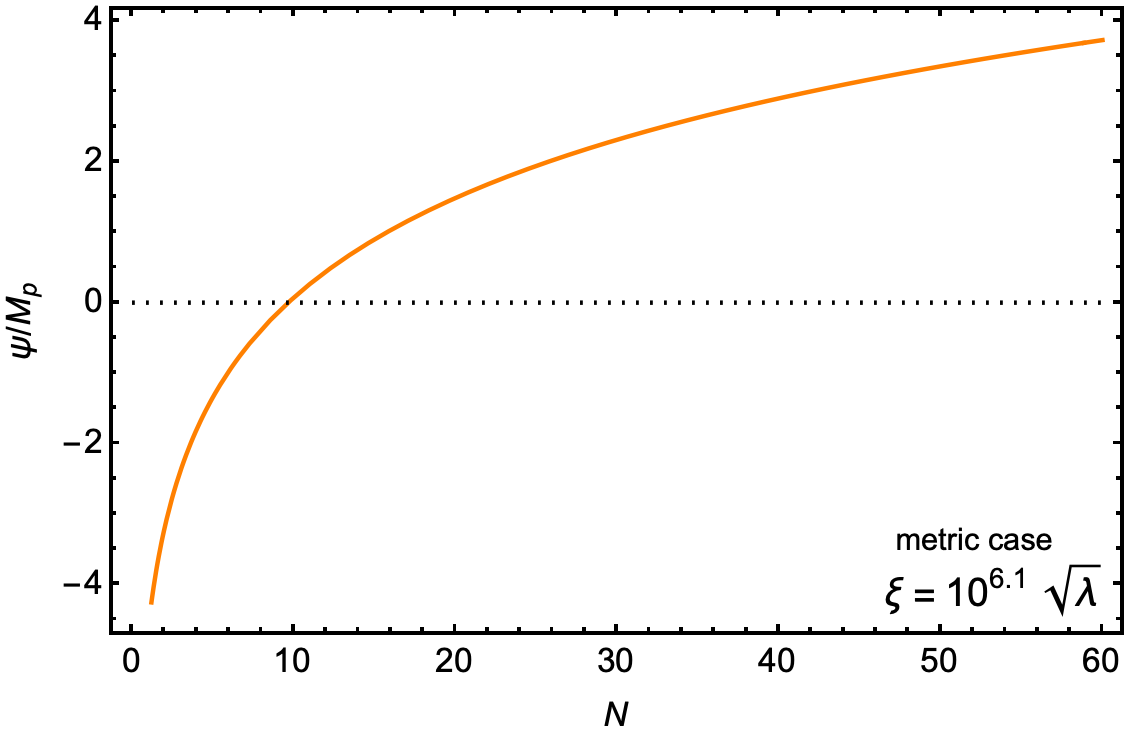}
 	\includegraphics[width=8cm]{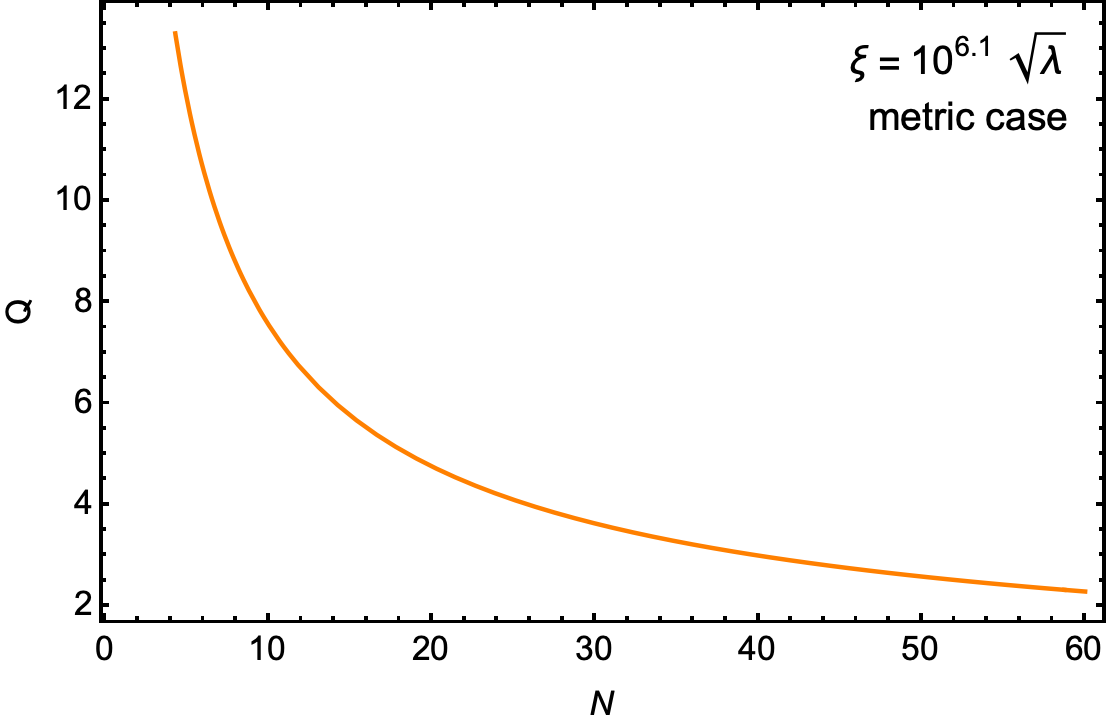}
  	\includegraphics[width=8cm]{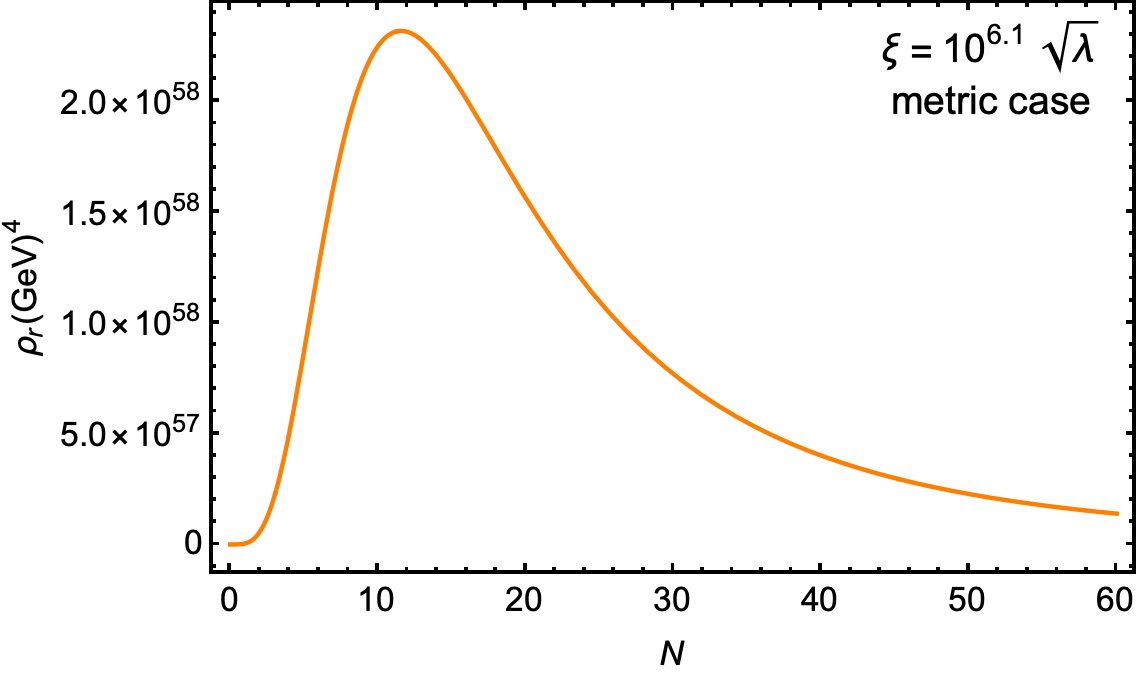}
   	\includegraphics[width=8cm]{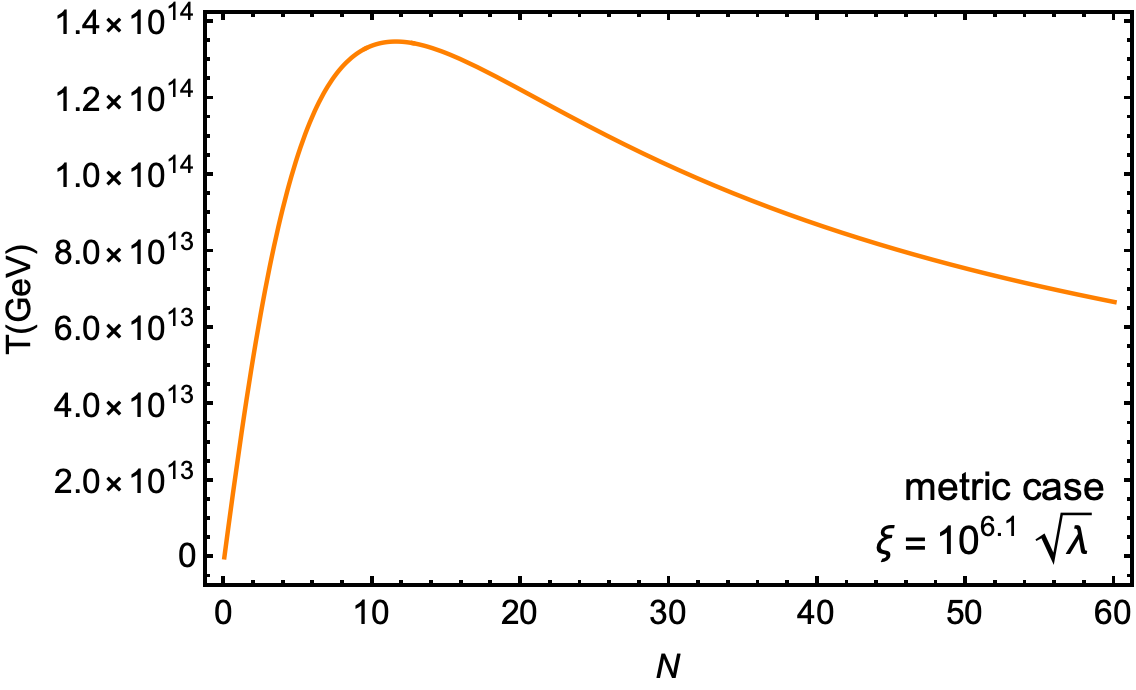}
	\centering
\caption{The behaviour of the homogeneous inflaton field $\psi$ (in units of $M_p$), the dissipation
parameter $Q$, the energy density in radiation, and temperature $T$ of the Universe is shown as a function of the number of efolds $N$ with the dissipation coefficient $\Gamma=C_{T}T$ in the metric case. To generate this plot, we take $\xi=10^{6.1}\sqrt{\lambda},\,C_{r}=70,\,C_{T}=0.045$.} \label{4expo}
\end{figure}

We show the behaviour of the evolution of $\psi$ (in units of $M_p$) and the temperature $T$ during warm inflation of the metric case in Fig.(\ref{4expo}). The dissipation parameter, $Q$, depending on both $\psi$ and $T$, is not a constant but rather evolves during inflation. This behaviour can also be seen in Fig.(\ref{4expo}). Additionally, as shown in Fig.(\ref{4expo}), we find that the energy density of radiation does not change appreciably when the modes of cosmological interest cross the horizon.

%%%%%%%%%%%%%%%%%%%%%%%%%%%%%%%%%
\subsection{Palatini Formalism}
%%%%%%%%%%%%%%%%%%%%%%%%%%%%%%%%
We follow the proceeding subsection. Since the energy density during inflation is predominated by the potential of the inflaton field, for the Palatini case, this allows us to write
\begin{eqnarray}
H^{2}=\frac{\lambda  M_{p}^2 \tanh ^4\left(\frac{\sqrt{\xi } \psi }{M_{p}}\right)}{12 \xi ^2}\,.\label{H2s}
\end{eqnarray}
Using the above relation, we can express Eq.(\ref{E1}) for this model as
\begin{eqnarray}
{\dot \psi}\approx -\frac{U'(\psi)}{3 (Q+1) H}=-\frac{\lambda  M_p^3 \tanh ^3\left(\frac{\sqrt{\xi } \psi }{M_p}\right) \text{sech}^2\left(\frac{\sqrt{\xi } \psi }{M_p}\right)}{2 \sqrt{3} \xi ^{3/2} (Q+1)}\left(\frac{C_{T}^4 \lambda  \xi  M_p^2}{\left(3 \sqrt{C_{r}} \sqrt{\lambda } Q^{5/2}+4 C_{T}^2 \xi ^{3/2}\right)^2}\right)^{-1/2}\,.\label{H2ps}
\end{eqnarray}
Using Eq.(\ref{H2s}) and Eq.(\ref{H2psis}), we come up with the following expression:
\begin{eqnarray}
\frac{H^{2}}{2\pi {\dot \psi}}=-\frac{(Q+1) \sinh \left(\frac{\sqrt{\xi } \phi }{M_{p}}\right) \cosh \left(\frac{\sqrt{\xi } \phi }{M_{p}}\right)}{12 \sqrt{3} \pi M_{p} \sqrt{\xi}}\left(\frac{C_{T}^4 M_{p}^2 \xi }{C_{r} Q^5 \left(\frac{4 C_{T}^2 \xi ^{3/2}}{3 \sqrt{C_{r}} \sqrt{\lambda } Q^{5/2}}+1\right)^2}\right)^{1/2}\,.\label{H2psis}
\end{eqnarray}
On substituting $Q=\Gamma/3H=C_{T}T/3H$ in the energy density of radiation given in Eq.(\ref{eomradsl}), we obtain the temperature of the thermal bath as
\begin{eqnarray}
T=\Bigg(\frac{\lambda M_{p}^4 Q \tanh ^2\left(\frac{\sqrt{\xi } \phi }{M_{p}}\right) \text{sech}^4\left(\frac{\sqrt{\xi } \phi }{M_{p}}\right)}{C_{r} \xi  (Q+1)^2}\Bigg)^{1/4}\,.\label{Ts}
\end{eqnarray}
We divide the above relation with $H$ to obtain
\begin{eqnarray}
\frac{T}{H}=\frac{9}{4 \sqrt{\frac{C_{T}^4 M_{p}^2 \xi }{C_{r} Q^5 \left(\frac{4 C_{T}^2 \xi ^{3/2}}{3 \sqrt{C_{r}} \sqrt{\lambda } Q^{5/2}}+1\right)^2}}}\left(\frac{\lambda ^2 M_{p}^4 Q^6 \tanh ^6\left(\frac{\sqrt{\xi } \phi }{M_{p}}\right) \text{sech}^4\left(\frac{\sqrt{\xi } \phi }{M_{p}}\right) \left(\frac{4 C_{T}^2 \xi ^{3/2}}{3 \sqrt{C_{r}} \sqrt{\lambda } Q^{5/2}}+1\right)^2}{C_{T}^4 \xi ^4 (Q+1)^2}\right)^{1/4}\,.\label{THs}
\end{eqnarray}
The dissipation parameter is defined as $Q=\Gamma/3H$. In this model of warm inflation, we have considered $\Gamma = C_{T}T$. On substituting this form of $\Gamma$ we get $T=3HQ/C_{T}$. We equate this with Eq.(\ref{Ts}) to obtain
\begin{eqnarray}
\frac{\psi}{M_p}=\frac{1}{\sqrt{\xi }}\sinh ^{-1}\left(\frac{2 C_{T} \xi ^{3/4}}{\sqrt{3} \sqrt[4]{C_{r}} \sqrt[4]{\lambda } Q^{3/4} \sqrt{Q+1}}\right)\,.\label{psis}
\end{eqnarray}

\begin{figure}[!h]	
	\includegraphics[width=8cm]{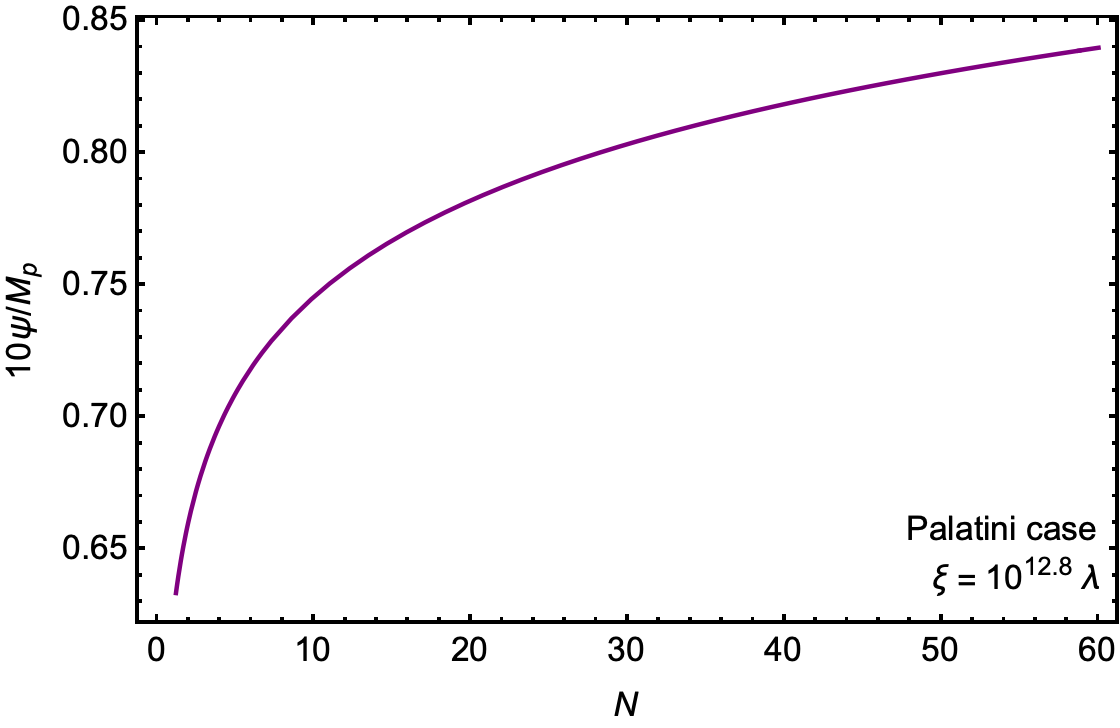}
 	\includegraphics[width=8cm]{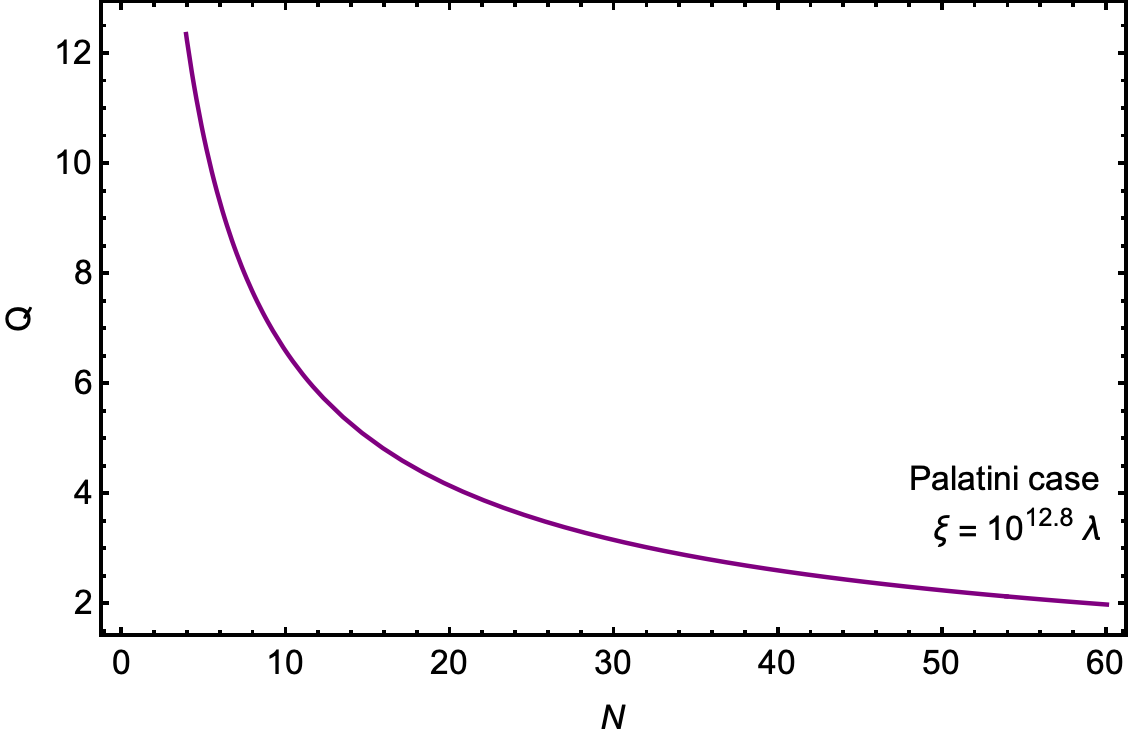}
  	\includegraphics[width=8cm]{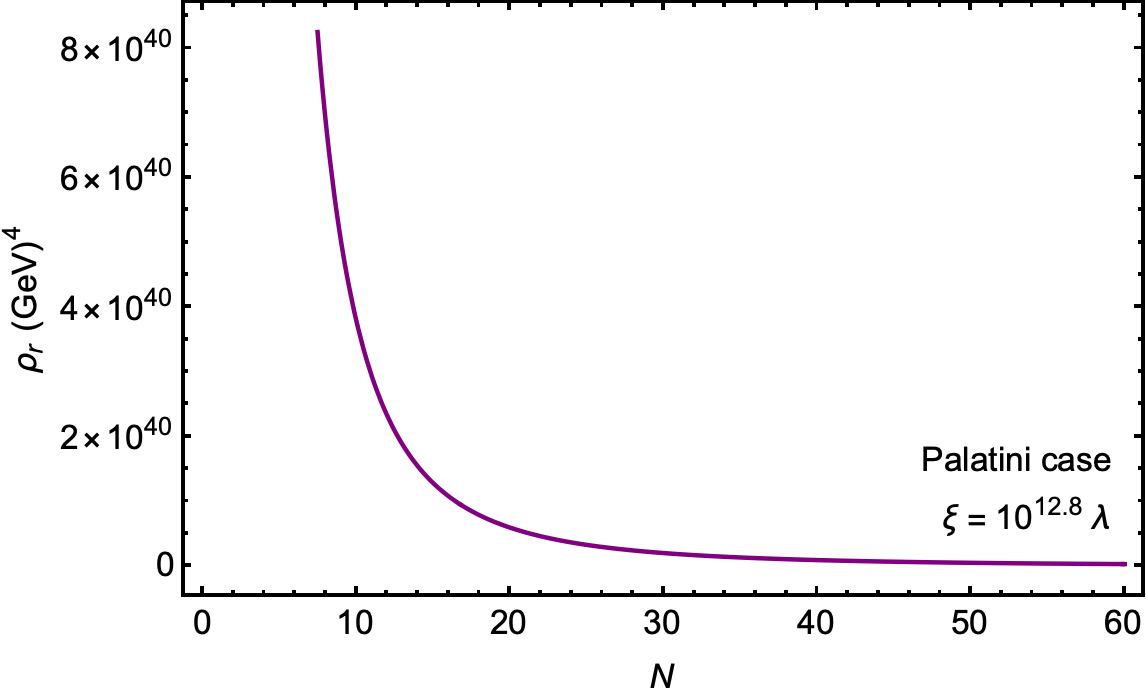}
   	\includegraphics[width=8cm]{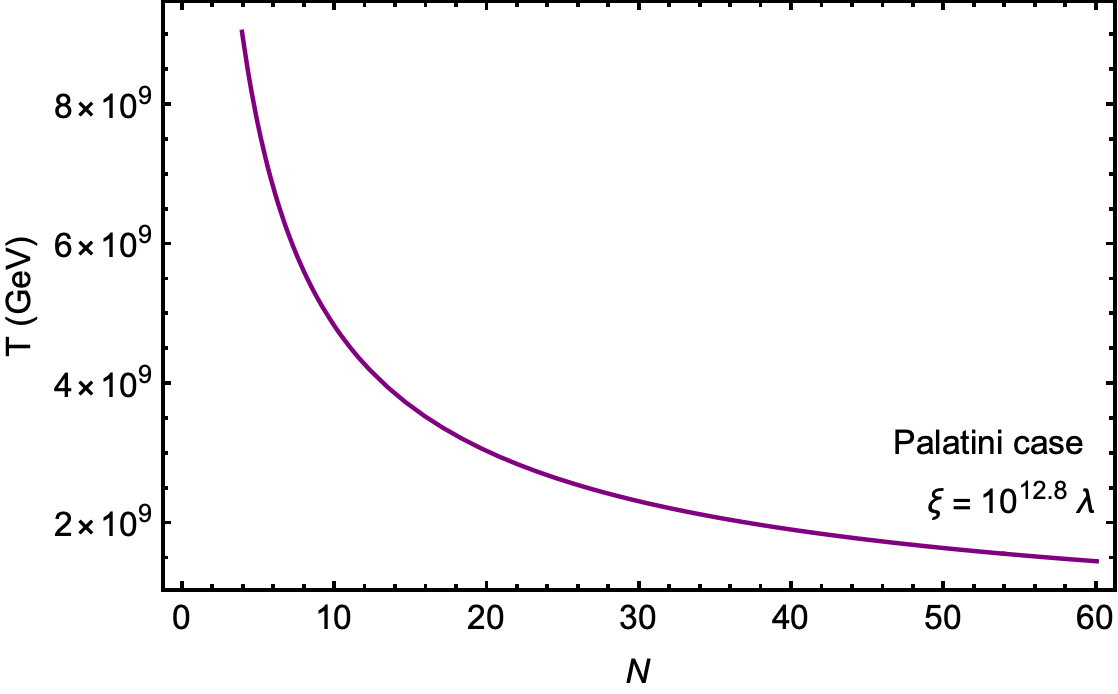}
	\centering
\caption{The behaviour of the homogeneous inflaton field $\psi$ (in units of $M_p$), the dissipation
parameter $Q$, the energy density in radiation, and temperature $T$ of the Universe is shown as a function of the number of efolds $N$ with the dissipation coefficient $\Gamma=C_{T}T$ in the Palatini case. To generate this plot, we take $\xi=10^{12.8}\lambda,\,C_{r}=70,\,C_{T}=0.045$.} \label{4Sinh}
\end{figure}

On substituting Eq.(\ref{psis}) in Eqs.(\ref{Ts}) and (\ref{H2psis}), we can express $P_{R}(k)$ in terms of variables $\xi,\,\lambda,\,Q$ and $C_T$. Also, from its definition in Eq.(\ref{slowroll}), the slow roll parameters can be written
\begin{eqnarray}
\varepsilon_{H}&=&\frac{8 \xi  \text{csch}^2\left(\frac{\sqrt{\xi } \psi }{M_{p}}\right) \text{sech}^2\left(\frac{\sqrt{\xi } \phi }{M_{p}}\right)}{Q+1}=\frac{6 \sqrt{C_{r}} \sqrt{\lambda } Q^{3/2}}{C_{T}^2 \sqrt{\xi } \left(\frac{4 C_{T}^2 \xi ^{3/2}}{3 \sqrt{C_{r}} \sqrt{\lambda } Q^{3/2} (Q+1)}+1\right)}\,,\\\eta_{H}&=&\frac{4 \xi ^2 \coth ^4\left(\frac{\sqrt{\xi } \phi }{M_{p}}\right)}{\lambda  M_{p}^2 (Q+1)}\Bigg(\frac{3 \lambda M_{p}^2 \tanh ^2\left(\frac{\sqrt{\xi } \phi }{M_{p}}\right) \text{sech}^4\left(\frac{\sqrt{\xi } \phi }{M_{p}}\right)}{\xi }-\frac{2 \lambda M_{p}^2 \tanh ^4\left(\frac{\sqrt{\xi } \phi }{M_{p}}\right) \text{sech}^2\left(\frac{\sqrt{\xi } \phi }{M_{p}}\right)}{\xi }\Bigg)\nonumber\\&=&\frac{9 C_{r} Q^5 \left(\frac{4 C_{T}^2 \xi ^{3/2}}{3 \sqrt{C_{r}} \sqrt{\lambda } Q^{5/2}}+1\right)^2}{4 C_{T}^4 M_{p}^2 \xi  (Q+1)}\Bigg(\frac{4 C_{T}^2 \sqrt{\lambda } M_{p}^2 \sqrt{\xi }}{\sqrt{C_{r}} Q^{5/2} \left(\frac{4 C_{T}^2 \xi ^{3/2}}{3 \sqrt{C_{r}} \sqrt{\lambda } Q^{5/2}}+1\right)^3}\nonumber\\&&\quad\quad\quad\quad\quad\quad\quad\quad\quad\quad\quad\quad\quad-\frac{32 C_{T}^4 M_{p}^2 \xi ^2}{9 C_{r} Q^5 \left(\frac{4 C_{T}^2 \xi ^{3/2}}{3 \sqrt{C_{r}} \sqrt{\lambda } Q^{5/2}}+1\right)^3}\Bigg)\,.
\end{eqnarray}
Using Eq.(\ref{H2}), the tensor power spectrum for this model is evaluated and we can use Eq.(\ref{psis}) and express $P_{T}(k)$ in terms of model parameters
\begin{eqnarray}
P_{T}(k)=\frac{16}{\pi}\Big(\frac{H}{M_{p}}\Big)^{2}=\frac{13824 \sqrt[3]{\frac{2}{5}} 3^{2/3} C_{T}^4 \xi  \left(\frac{\sqrt{C_{r}} \sqrt{\lambda } n}{C_{T}^2 \sqrt{\xi }}\right)^{10/3}}{125 \pi  C_{r} \left(\frac{72 \sqrt[3]{3} \left(\frac{2}{5}\right)^{2/3} C_{t}^2 \xi ^{3/2} \left(\frac{\sqrt{C_{r}} \sqrt{\lambda } n}{C_{T}^2 \sqrt{\xi }}\right)^{5/3}}{5 \sqrt{C_{r}} \sqrt{\lambda }}+1\right)^2}\,.
\end{eqnarray}
In this subsection, we will evaluate how the dissipation parameter, $Q$, evolves with the number of efolds, $N$. We differentiate Eq.(\ref{psis}) w.r.t $N$ and then again write $d\psi/dN=-{\dot\psi}/H$. By using Eqs. (\ref{H2s}), (\ref{H2ps}) and (\ref{psis}), we obtain
\begin{eqnarray}
\frac{dQ}{dN}=-\frac{12 \sqrt{C_{r}} \sqrt{\lambda } Q^{5/2}}{5 C_{T}^2 \sqrt{\xi }}\,,
\end{eqnarray}
where we have assumed a large field approximation $\xi\phi^{2}/M_{p}^{2}\gg 1$ or equivalently $\psi\gg\sqrt{3/2}M_{p}$.

The behaviour of the evolution of $\psi$ (in units of $M_p$) and the temperature $T$ during warm inflation of the Palatini case is displayed in Fig.(\ref{4Sinh}). Similarly, the dissipation parameter, $Q$, depending on both $\psi$ and $T$, is also not a constant but rather evolves during inflation. This behaviour can also be seen in Fig.(\ref{4Sinh}). We also find that the energy density of radiation does not change appreciably when the modes of cosmological interest cross the horizon shown in Fig.(\ref{4Sinh}).

%%%%%%%%%%%%%%%%%%%%%%%
\section{Confrontation with the Planck 2018 data}\label{C5}
\begin{figure}[!h]	
	\includegraphics[width=8cm]{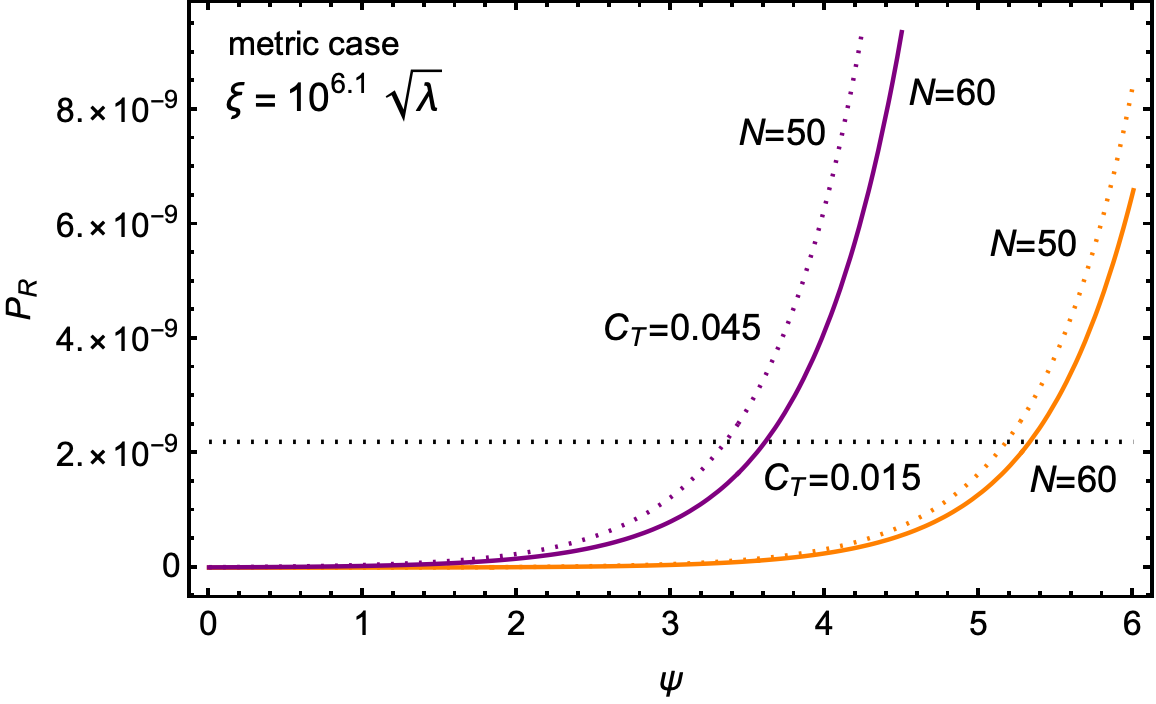}
 	\includegraphics[width=8cm]{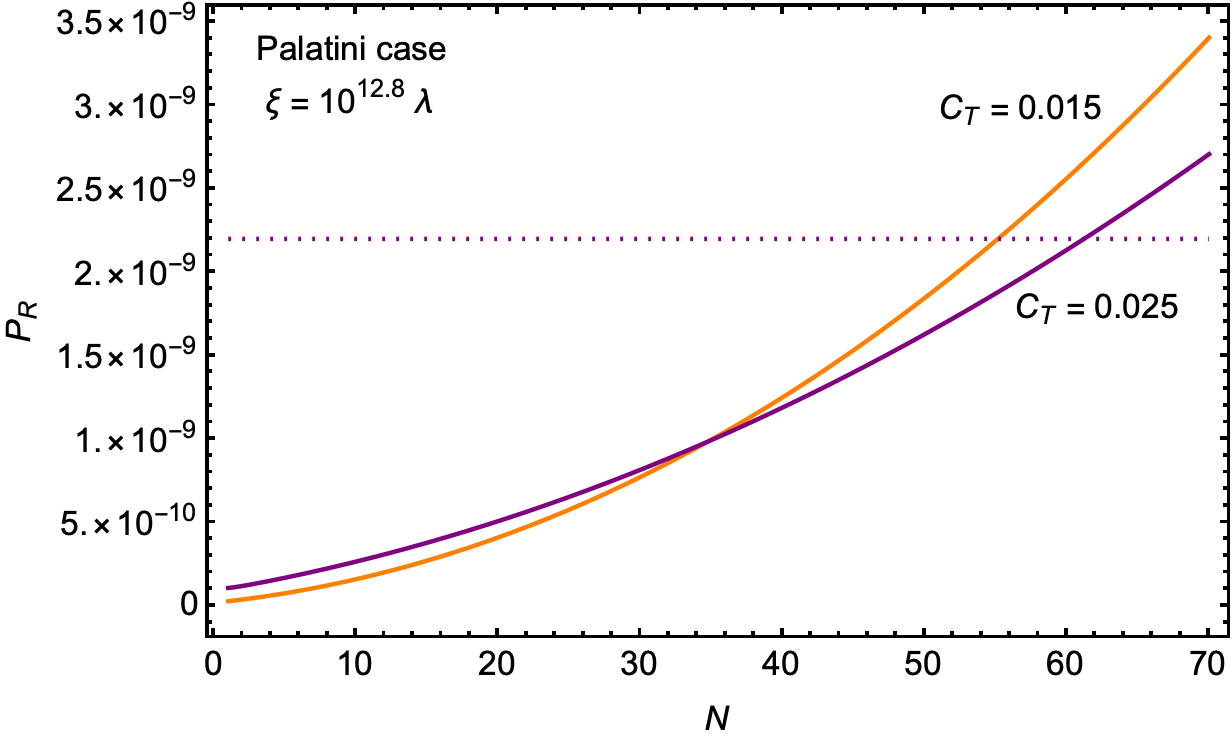}
	\centering
	\caption{We constrain our models with the amplitude of the primordial power spectrum. Left panel, we showed the metric case with $\xi\sim 1.26\times 10^{6}\sqrt{\lambda}$ and $C_{T}=0.015,\,0.045$ with $N=50,\,60$. The parameters used provide the preferred amplitude of the primordial power spectrum of $A_{s}\sim2.2 \times 10^{-9}$. Right panel: we displayed the Palatini case with $\xi\sim 6.31\times 10^{12}\lambda$ and $C_{T}=0.015,\,0.025$. The parameters used also provide the preferred amplitude of the primordial power spectrum. A dotted horizon line denotes $A_{s}=2.2 \times 10^{-9}$.}
	\label{sPR}
\end{figure}

%%%%%%%%%%%%%%%%%%%%%%%
We constrain our results using the amplitude of the primordial power spectrum. Consider Eq.(\ref{spectrum}) we find that our predictions can produce the prefered values of $P_{R}\sim A_{s}=2.2 \times 10^{-9}$ shown in Fig.(\ref{sPR}). We notice for the metric case that in order to produce a corrected value of $P_{R}$, when we decrease values of $C_{T}$, the magnitudes of $\psi$ get increased. However, in the Palatini case, when we decrease values of $C_{T}$, a number of efolds get decreased.

\begin{figure}[!h]	
	\includegraphics[width=8cm]{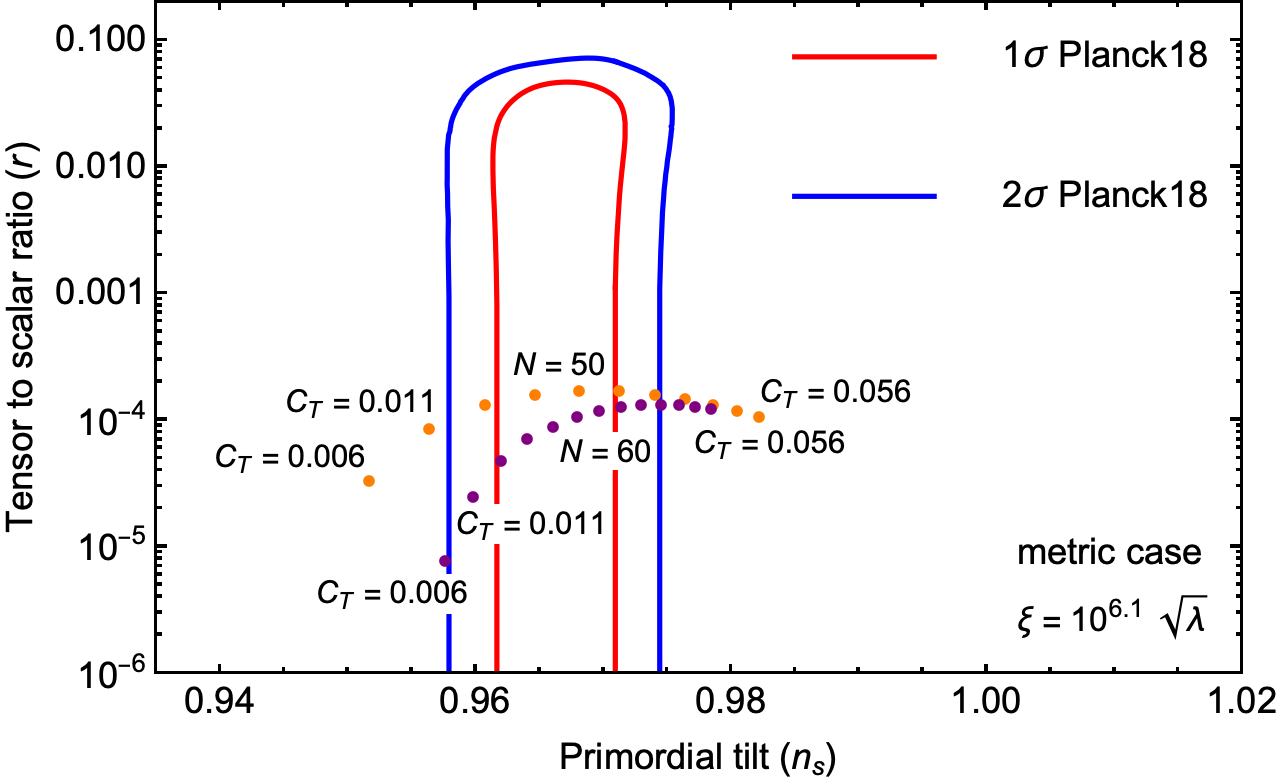}
 	\includegraphics[width=8cm]{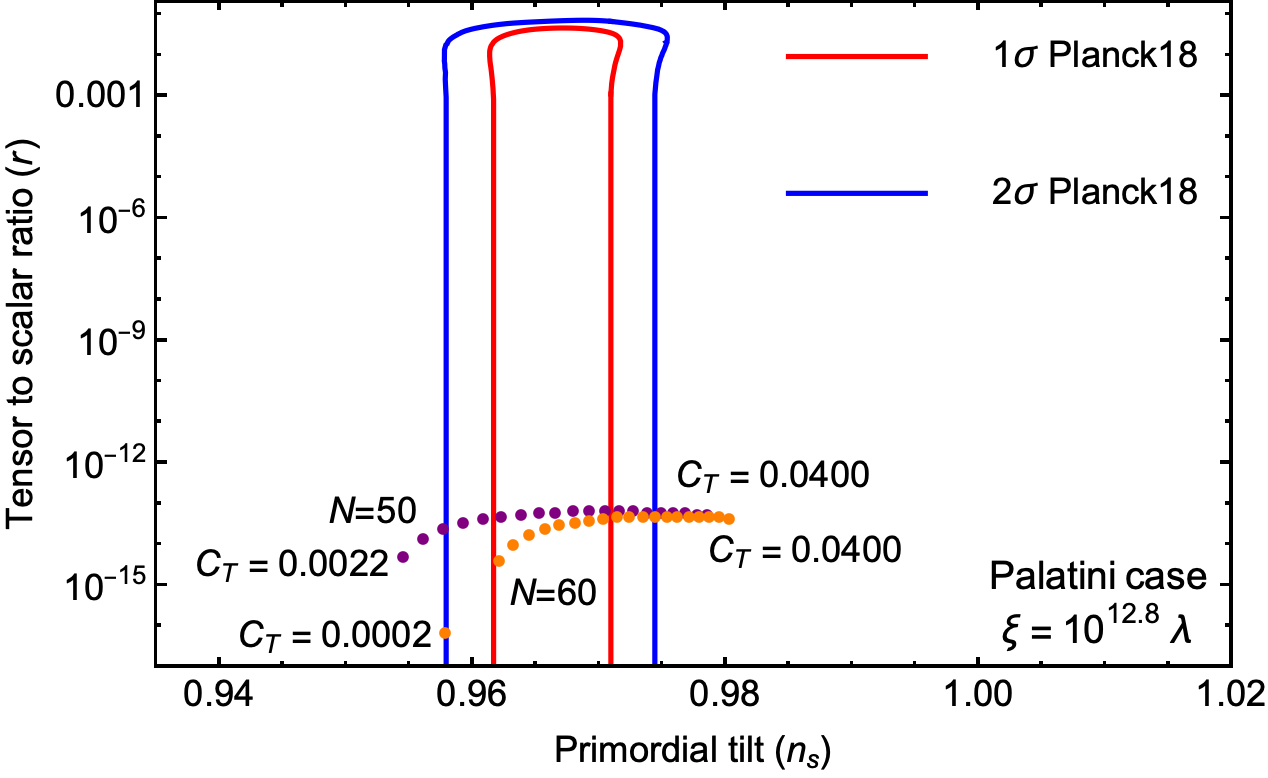}
	\centering
	\caption{We compare the theoretical predictions of $(r,\,n_s)$ in the strong limit $Q\gg 1$ for the metric (left panel) and Palatini (right panel) approaches. We consider a linear form of the growing mode function $G(Q_N)$. For the plots, we have used $C_{r}=70,\,\xi=1.26\times 10^{6}\sqrt{\lambda}$ for the metric case, and $C_{r}=70,\,\xi=6.31\times 10^{12}\lambda$ for the Palatini case. We consider theoretical predictions of $(r,\,n_s)$ for different values
of $C_{T}$ with Planck’18 results for TT, TE, EE, +lowE+lensing+BK15+BAO.}
	\label{nsr2}
\end{figure}

We compute the inflationary observables and then compare with the Plank 2018 data. We plot the derived $n_s$ and $r$ for our models along with the observational constraints from Planck 2018 data displayed in Fig.(\ref{nsr2}). Left panel, we used $\xi=10^{6.1}\sqrt{\lambda},\,C_{r}=70$ and $N=50,\,60$ for $C_{T}\in [0.001, 0.06]$. Our results obtained in the metric case show that, for $N=50$, $C_{T}\in [0.0161,0.0276]$, while for $N=60$, $C_{T}\in [0.008,0.0206]$ is required in order to have the derived $n_{s}$ consistent with the Planck 2018 observations at $1\sigma$ CL. Additionally, we can obtain $n_{s}=0.9649$ using $C_{T}=0.0214$ and $C_{T}=0.0139$ for $N=50$ and $N=60$, respectively. 

Likewise, for the right panel, we used $\xi=10^{12.8}\lambda,\,C_{r}=70$ and $N=50,\,60$ for $C_{T}\in [0.001, 0.06]$. Our results obtained in the metric case show that, for $N=50$, $C_{T}\in [0.010,0.022]$, while for $N=60$, $C_{T}\in [0.00053,0.0141]$ is required in order to have the derived $n_{s}$ consistent with the Planck 2018 observations at $1\sigma$ CL. Additionally, we can obtain $n_{s}=0.9649$ using $C_{T}=0.0156$ and $C_{T}=0.007$ for $N=50$ and $N=60$, respectively.

%%%%%%%%%%%%%%%%%%%%%%%
\section{Conclusion}
%%%%%%%%%%%%%%%%%%%%%%%

In this work, we studied the non-minimally-coupled Higgs model in the context of warm inflation scenario using both metric and Palatini approaches. We particularly considered a dissipation parameter of the form $\Gamma=C_{T}T$ with $C_{T}$ being a coupling parameter and focused only on the strong regime of the interaction between inflaton and radiation fluid. We compute all relevant cosmological parameters and constrained the models using the observational Planck 2018 data. We discovered that the $n_s$ and $r$ values are consistent with the observational bounds. Having used the observational data, we obtained a relation between $\xi$ and $\lambda$ for the non-minimally-coupled warm Higgs inflation in both metric and Palatini cases. Our constraints on the parameters are compatible with Planck data. Furthermore in comparison to other literature on the topic,  we proposed that the computed $(r,\,n_s)$ parameters values are two orders of magnitude higher than those of the usual (cold) non-minimally-coupled Higgs inflation.

Having compared between two approaches, the energy density and the temperature of the thermal bath in the metric case, see Fig.\ref{4expo}, are many orders of magnitude larger than those found in the Palatini case, see Fig.\ref{4Sinh}. To produce $n_s$ and $r$ in agreement with observation, we found that their values are two orders of magnitude higher than those of the usual (cold) non-minimally-coupled Higgs inflation \cite{Bauer:2008zj,Jinno:2019und}. However, we noticed that the ratio of $\xi^{2}/\lambda$ of the metric case in this work are four orders of magnitude higher than that of model present in Ref.\cite{Kamali:2018ylz}. This may can quantify the amount of primordial gravitational waves produced during the inflationary epoch between cold and warm Higgs inflation. Since the value of $r$ depends on the specific inflationary model, different models predict different amounts of gravitational waves generated during inflation. A lower value of $r$ implies weaker gravitational waves, while a higher value indicates stronger gravitational waves.

It is worth mentioning that in standard inflationary models, the inflaton field is minimally coupled to gravity, meaning its dynamics are governed solely by the Einstein equations. However, in warm inflation, a non-minimal coupling term of the form $\xi H^{2} R$ is introduced, where $\xi$ is the coupling constant, $H$ is the inflaton field, and $R$ is the scalar curvature. In the context of warm inflation, where there is dissipative particle production and energy transfer between the inflaton and other fields, the non-minimal coupling can influence the dissipation mechanism. The coupling term introduces additional interactions between the inflaton and the thermal bath of particles, affecting the dissipation coefficient and the energy transfer rate. In suumary, the effects of the non-minimal coupling on the dissipative term in warm inflation can influence the energy transfer, particle production, backreaction effects, and stability of the inflationary dynamics. These effects play a significant role in determining the observational predictions and the viability of warm inflation models. We will leave these interesting issues for our future investigation.

\acknowledgments
P. Channuie acknowledged the Mid-Career Research Grant 2020 from National Research Council of Thailand (NRCT5-RSA63019-03).

%%%%%%%%%%%%%%%%%%%%%%%%%%%%%%%%%%%%%%%%
%%%%%%%%%%%%%%%%%%%%%%%%%%%%%%%%%%%%%%%%
\end{document}